\documentstyle[a4wide,epsfig]{article}

\newcommand{\AmS}{{\protect\the\textfont2
  A\kern-.1667em\lower.5ex\hbox{M}\kern-.125emS}}
\newcommand{\ltsim}{\lower3pt\hbox{$\, \buildrel < \over \sim \, $}}
\newcommand{\gtsim}{\lower3pt\hbox{$\, \buildrel > \over \sim \, $}}

\begin{document}

\rightline{UG--FT--143/02}
\rightline{CAFPE--13/02}
\rightline{IPPP/02/68}
\rightline{hep-ph/0212205}
\begin{center}

\vspace{0.5cm}

\large {\bf 
Low energy constraints on orbifold models
\footnote{Presented at RADCOR 2002, Kloster Banz, Germany, September 8-13,
  2002, to appear in the proceedings. Work partially
supported by the EU (HPRN-CT-2000-00149), MCYT (FPA2000-1558), Junta
de Andaluc\'\i a (group FQM 101) and PPARC.}
}
\vspace*{5mm}
\normalsize

{\bf F. del Aguila$^a$ and J. Santiago$^{a,b}$}

\smallskip 
\medskip 
$^a${\it Departamento de F\'\i sica Te\'orica y del Cosmos and 
Centro Andaluz de F\'\i sica de \\Part\'\i culas Elementales}
(CAFPE),
{\it Universidad de Granada, E-18071 Granada, Spain}
\\
\medskip 
$^b${\it  
Institute of Particle Physics
                Phenomenology, 
University of Durham, Durham, DH1 3LE, UK}
\smallskip  

\vskip0.6in \end{center}

\centerline{\large\bf Abstract}
\noindent
We review the low energy limits on Kaluza-Klein excitations 
in orbifold models. New vector-like quarks, as well as new 
$Z'$ gauge bosons, can be accommodated with masses observable 
at large colliders. 

\section{INTRODUCTION}

Although the Standard Model (SM) is in agreement with experiment up to
energies near the electroweak scale~\cite{Hagiwara:pw}, 
it is believed to be an effective
theory. Many of its completions, as for instance string theory,
require extra dimensions which, if large enough, may manifest at large
colliders.

Let $M_c$ be the compactification scale with $R\sim M_c^{-1}$ the size
of the largest extra dimension. If the scale $M_s$ of the more
fundamental theory is larger than $M_c$, there should be a range of
energies where physics can be described by an effective theory in
$d>4$ dimensions. Such theories are interesting not only by their
phenomenological implications but for giving new insights into more
theoretical questions like the replica of families or the breaking of
symmetries. All these questions have received a great deal of attention.
Here we want to review the prospects for having new matter and/or
interactions at the TeV scale, ignoring at this stage new
gravitational effects.

Even in the case of one extra dimension, the simplest models allow for
a rich phenomenology. Fields living in the fifth dimension, bulk
fields, can be described by an infinite tower of Kaluza-Klein (KK)
modes. The known particles corresponding to the zero modes and the
heavy excitations to new exotic fermions and/or gauge bosons and
scalars. Typically the heavy masses scale with $M_c$ which plays the
role of the order parameter of the physics beyond the
SM. Thus models can be classified according to the
present experimental 
bounds on its value. In Table~\ref{tabla} we order the different
classes of models with growing compactification scale limits. 

\begin{table*}[!bht]
\caption{Classification of models with extra dimensions according to
  the present experimental bounds on the compactification scale $M_c$.
\label{tabla}}
\begin{tabular}{lcc}
Class of models & $M_c$ bound & Constraining observable \\
\hline
\hline
Universal
& $M_c \gtsim$  300 GeV & 
T parameter~\cite{Appelquist:2000nn}
\\
\hline
Brane interactions 
& $M_c \gtsim$ 4 TeV & 
Precision data~\cite{Masip:1999mk,Delgado:1999sv} 
\\
\hline
Randall-Sundrum
& $M_c \gtsim$  10 TeV & 
Precision data~\cite{Huber:2000fh}
 \\
\hline
Fermion spreading 
& $M_c \gtsim$  100 TeV & 
FCNC~\cite{Delgado:1999sv,Carone:1999nz}
\\
\hline
Fermion
multilocalization
 & \begin{tabular}{c} $M_c \gtsim$  80 TeV \\
  ${\mathrm{e}}^{-M/M_c} M_c \sim$ 300 GeV \end{tabular} & 
\begin{tabular}{c}Precision and \\
production data~\cite{DelAguila:2001pu}\end{tabular}
\\ \hline 
\end{tabular}
\end{table*}


            As observed in Ref.~\cite{Appelquist:2000nn}, 
if there is conservation of the
KK number, there are no couplings of SM fields with
only one heavy mode. Hence the new contributions to SM
observables come only at one loop, implying an extra suppression
factor $\sim \alpha$. This translates into a relatively low
$M_c$ limit, $\sim 300$ GeV, mainly resulting from the T parameter
and the running of the heavy excitations around the loop.
A simple realization of this scenario is a five-dimensional
theory with the extra dimension parametrizing an orbifold
$S^1/Z_2$: $-\pi R \leq y \leq \pi R$ with opposite points
identified, and all fields in the bulk and without brane terms.

            The addition of brane interactions violates
momentum conservation in the fifth dimension and introduces
new vertices involving SM particles and only one heavy mode.
Then, if standard matter is assumed to live only in the brane
in order to avoid large Flavour Changing Neutral Currents (FCNC), the
strongest constraint on the compactification scale results
from the tree level exchange of the KK tower of
gauge bosons, implying $M_c \gtsim 4$ TeV to maintain the
electroweak parameters within experimental
limits~\cite{Masip:1999mk,Delgado:1999sv}.

            These simple models have a flat background but
enough physics to share the main new effects. Nature can
choose, however, a more general situation with a non-trivial
background, as for instance in the case 
of the
Randall-Sundrum model~\cite{Randall:1999ee}. 
A detailed analysis gives in this case
$M_c \gtsim 10$ TeV~\cite{Huber:2000fh}.

            A different scenario appears when the models
are generalized to include fermions in the bulk in order
to explain the observed pattern of fermion masses and
mixings geometrically, split fermions~\cite{Arkani-Hamed:1999dc}. 
Doing so the
disalignment of the KK gauge couplings to
SM fermions generates large FCNC~\cite{Delgado:1999sv,Carone:1999nz}, 
which banish the
gauge boson masses and $M_c$ to high energy, $\gtsim 100$ TeV,
making them unobservable at future colliders. This is
only a particular case. When general five-dimensional
masses and/or brane kinetic terms are introduced, the
masses of the lightest KK excitations can decouple
from $M_c$ due to their dependence on the size of
the new terms, and become observable.
Here we concentrate
on this type of models and their manifestation at large
colliders.

In the following we review the KK formulation of these models. 
Then we first introduce five-dimensional mass terms for the
fermions and discuss simple models with multi-localization
of the lightest KK fermion wave function in the
extra dimension. In this way the observed
pattern of fermions masses and mixings can be obtained with order one
Yukawa couplings, together with relatively light exotic
fermions~\cite{delAguila:2000rc} 
decoupled from a rather heavy compactification scale
and observable at large colliders~\cite{DelAguila:2001pu,thesis}. 
Afterwards,
we consider brane kinetic terms for gauge bosons~\cite{Carena:2002me}. For a
choice of terms one can also reduce the mass of the
lightest KK gauge boson excitation, making
it eventually observable. This can also be done for
fermions. In both cases the decoupling of the lightest
heavy masses are not due to the multi-localization of
the wave functions but to a sharper localization.

\section{FOUR-DIMENSIONAL FIELDS, MASSES AND COUPLINGS}
An effective five-dimensional theory whose Lagrangian reads
\begin{equation}
\mathcal{L}(\phi)=\mathcal{L}_{\mathrm{kin}}(\phi)+
\mathcal{L}_{\mathrm{int}}(\phi), 
\end{equation}
can be described in terms of four-dimensional fields decomposing the
corresponding five-dimensional fields in four-dimensional
(KK) modes
\begin{equation}
\phi(x,y)=\sum_n 
\phi^{(n)}(x)
\frac{f_n(y)}{\sqrt{2\pi R}}.
\end{equation}
After integrating the extra dimension one is left with a
four-dimensional Lagrangian with an infinite number of KK
fields $\phi^{(n)}(x)$ with masses $m_n$, gauge couplings $g_{n\ldots}$
and Yukawa couplings $\lambda_{n\ldots}$. In this procedure the wave
functions $f_n(y)$ are fixed requiring orthonormality conditions, in order to
guarantee canonical kinetic terms for  $\phi^{(n)}$, and fulfilling the
differential equations giving them diagonal masses. Obviously
different ${\mathcal{L}}_{\mathrm{kin}}$ 
give different 
wave functions, thus modifying the
gauge and Yukawa couplings, and the physics derived from them.
For a bulk gauge boson $A_\mu $ and a fermion $\Psi $
\begin{equation}
{\mathcal{L}}_{\mathrm{gauge}}=
-\int {\mathrm{d}}y\; g_5 \bar{\Psi} \gamma^\mu A_\mu \Psi,
\end{equation}
and after decomposing and integrating the fifth dimension
\begin{equation}
{\mathcal{L}}_{\mathrm{gauge}}=
- \sum_{nmr} \left[
g^{LLV}_{nmr} \bar{\Psi}^{(n)}_L \gamma^\mu A_\mu^{(r)} \Psi_L^{(m)} +
(L\to R) \right],
\end{equation}
with
\begin{equation}
g^{LLV}_{nmr}
=\frac{g_5}{\sqrt{2\pi R}} \int_{-\pi R}^{\pi R} {\mathrm{d}}y\;
\frac{ f^L_n f^L_m f^V_r}{2\pi R} 
\label{4Dgauge}
\end{equation}
the dimensionless four-dimensional gauge couplings.
Similarly,
\begin{eqnarray}
\mathcal{L}_{\mathrm{Yuk}}&=&
-\int {\mathrm{d}}y\; \left[ \lambda_5 \bar{\Psi} \chi \phi +
  {\mathrm{h.c.}}\right] \nonumber \\
&=& - \sum_{nmr} \left[
\lambda^{\Psi \chi S}_{nmr} \bar{\Psi}^{(n)}_L \chi^{(m)}_R \phi^{(r)} 
\right.
\left.+
\lambda^{\chi \Psi S}_{mnr} \bar{\Psi}^{(n)}_R \chi^{(m)}_L \phi^{(r)} 
+{\mathrm{h.c.}}\right],
\end{eqnarray}
with
\begin{equation}
\lambda^{\Psi_1 \Psi_2 S}_{nmr}
=
\frac{\lambda_5}{\sqrt{\pi R}} \int_{-\pi R}^{\pi R} {\mathrm{d}}y\;
\frac{ f^{\Psi_1 L}_n f^{\Psi_2 R}_m f^S_r}{2\pi R} 
\label{4DYukawa}
\end{equation}
the dimensionless four-dimensional Yukawa couplings.
Tree level estimates of the contribution of the KK
excitations are derived in the properly normalized mass
eigenstate basis using $g$ and $\lambda $ in Eqs.~(\ref{4Dgauge}) and
(\ref{4DYukawa}),
respectively. For the particular case of a boundary Higgs
(without scalar KK tower)
\begin{equation}
\lambda_{nm}
=\frac{\tilde{\lambda}_5}{2\pi R} f^{\Psi L}_n(0) f^{\chi
  R}_m(0),
\end{equation}
where the tilde stands for the different dimension
normalization.


\section{FIVE-DIMENSIONAL FERMION MASSES AND MULTILOCALIZATION}

\begin{figure}[h]
\begin{center}
\vspace{-.5cm}
\includegraphics[width=7cm]{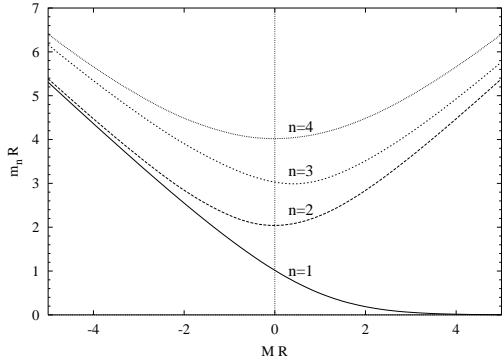}
\end{center}
\vspace{-.5cm}
\caption{Values of the masses of the first KK modes as a function of
  the five-dimensional mass in units of $1/R$. We have fixed $a=R/2$
in Eq.~(\ref{masa}).\label{spectrum}}
\end{figure}

\begin{figure}[h]
\begin{center}
\vspace{-.5cm}
\includegraphics[width=7cm]{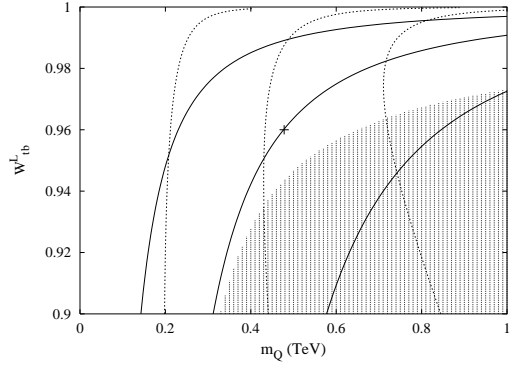}
\vspace{-.5cm}
\end{center}
\caption{Allowed region for the top coupling $W^L_{tb}$ and the lightest
  vector-like quark mass $m_Q$ as a function of the parameters of the
  model: solid (dotted) lines correspond to constant $a$ ($M$).
\label{goniometry}}
\end{figure}

  Models with extra dimensions can explain the observed
hierarchy of fermion masses and mixings relating them to the
localization of the corresponding zero modes around different
points in the extra dimensions. However, the lightest modes 
are often not localized but multi-localized if the
five-dimensional fermion masses are non-trivial.
Indeed, the addition of a step-function mass term in the
$S^1/Z_2$ orbifold case allows for the bi-localization
of the first two fermion modes. 
As shown in Ref.~\cite{DelAguila:2001pu} for
\begin{equation}
{\mathcal{L}}_{\mathrm{kin}}=\int {\mathrm{d}}y\;
\bar{\Psi} [ {\mathrm{i}}\gamma^N \partial_N -M(y)] \Psi,
\end{equation}
with $\gamma^N=\{\gamma^\mu,{\mathrm{i}}\gamma^5\}$ and
\begin{equation}
M(y)=M \sigma(y)\sigma(\pi a-|y|), \quad \sigma(y)\equiv |y|/y,
\label{masa}
\end{equation}
one obtains the mass spectrum in Fig.~\ref{spectrum}
solving
the corresponding normalization and eigenvalue equations.
\begin{figure}[h]
\begin{center}
\includegraphics[width=4.5cm]{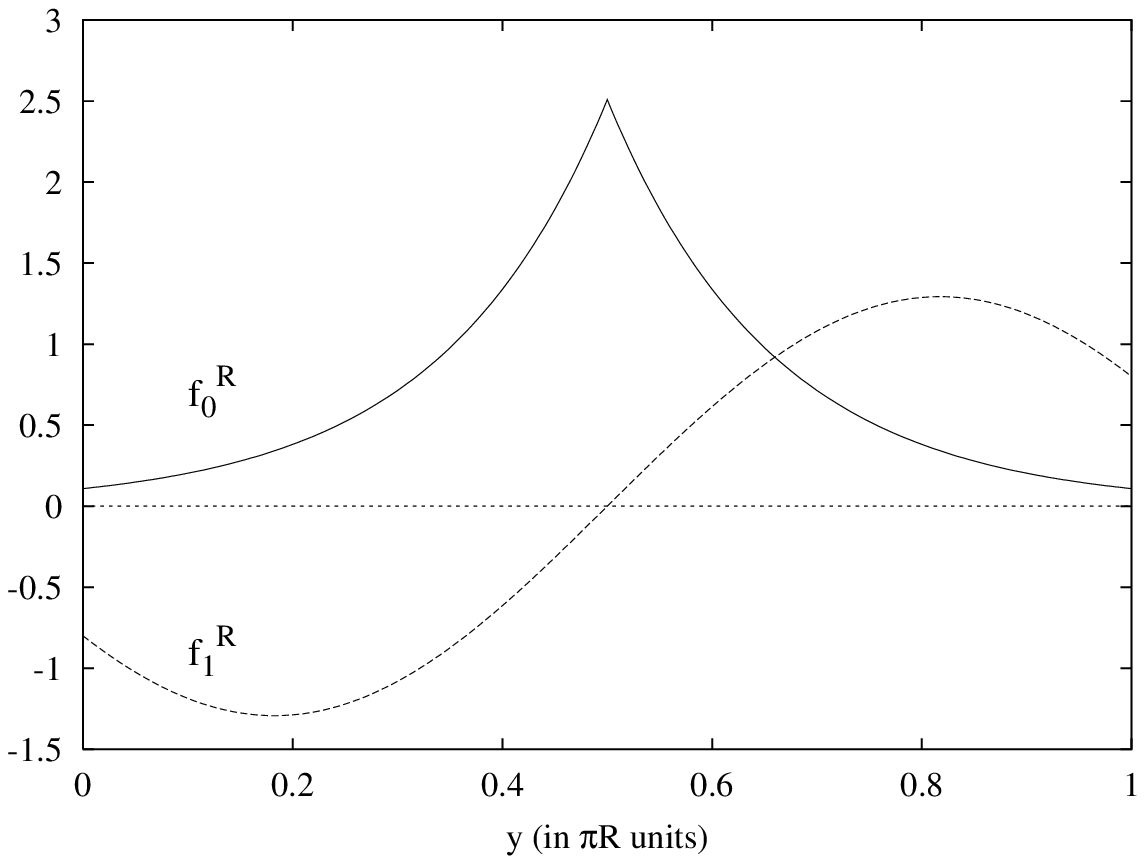}
\hspace{-0.25cm}
\includegraphics[width=4.5cm]{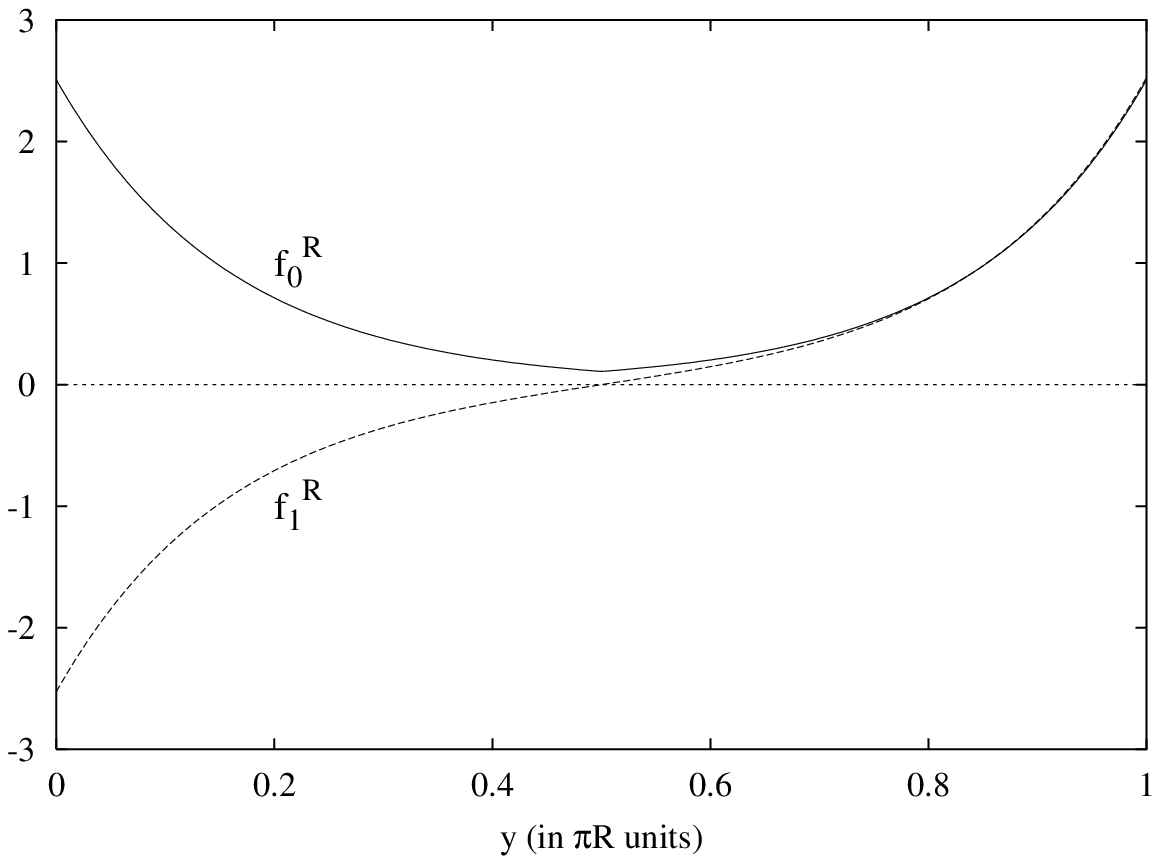}
\vspace{-.5cm}
\end{center}
\caption{Profiles of the massless zero mode $f_0^R$ and the first KK
  excitation $f_1^R$ with no multi-localization, $MR=-2$ (left), and
  with multi-localization, $MR=2$ (right).\label{wavefunctions}}
\end{figure}
The corresponding wave functions for the zero mode and the
lightest KK excitation are drawn in
Fig.~\ref{wavefunctions}. (We choose the Right-Handed (RH) component even.)
For $M=0$ one obtains the usual equidistant spectrum, whereas
for negative $M$ the four-dimensional masses get extra larger
contributions but for positive $M$ the lightest KK
mode can be as light as required. In this case the difference
between the absolute values of the two wave functions in
Fig.~\ref{wavefunctions} are exponentially suppressed. Many (split fermion)
models give a similar behaviour~\cite{Kaplan:2001ga}.
This type of models can accommodate in a natural way the observed
spectrum of masses and mixings (starting with five-dimensional Yukawa
couplings of order one), and at the same time an observable departure
from the SM at large colliders~\cite{DelAguila:2001pu,thesis}. 
In Fig.~\ref{goniometry} we plot the predicted mass of a new quark of
charge $2/3$ and the third family charged current coupling as a
function of the parameters of the model~\footnote{Multi-localization
  can also have interesting phenomenological implications for neutrino
  oscillations~\cite{Branco:2002hc}.}. The shaded region is
excluded by electroweak precision data.


\section{BRANE KINETIC TERMS AND LIGHT NEW PARTICLES}

Brane kinetic terms are naturally present in
orbifold theories~\cite{Georgi:2000ks}. They cannot
be set to zero at all scales in general, since the breaking of translation
invariance induces them at the quantum level. In this case the
equations of motion are modified leading to a different spectrum which
can include light KK states. In this section we review examples in
which the inclusion of brane kinetic terms can induce a light first KK
mode for gauge bosons and/or fermions. We do not intend to be exhaustive
here but just show, with the very practical goal of getting light KK
excitations, the effects of some of these terms and discuss the
corresponding phenomenology. A more detailed study will be
presented elsewhere~\cite{apvs}.

Let us consider the Lagrangian for a five-dimensional ($U(1)$
for simplicity) gauge theory with the following brane kinetic
terms~\cite{Carena:2002me} 
\begin{equation}
{\mathcal{L}}
=
-\frac{1}{4}
\int {\mathrm{d}}y\; \Big\{ F^{MN}F_{MN} + \big[a_0 \delta_0 + a_\pi
  \delta_\pi \big] F^{\mu\nu}F_{\mu \nu}
\Big\},
\end{equation}
where $\delta_0\equiv \delta(y)$ and $\delta_\pi\equiv \delta(y-\pi R)$. 
Performing the KK expansion in the Lagrangian and requiring the
quadratic terms to decouple, we arrive at the following orthonormality
and differential equation conditions
\begin{equation}
\int {\mathrm{d}}y\; [1+a_0 \delta_0+a_\pi \delta_\pi] \frac{f_m
  f_n}{2\pi R}= \delta_{nm}, 
\end{equation}
\begin{equation}
\partial_y^2 f_n(y)=-m_n^2(1+a_0 \delta_0 +a_\pi \delta_\pi) f_n(y),
\end{equation}
respectively.
The
corresponding spectrum consists on a massless flat zero mode plus a
tower of massive modes with wave functions
\begin{equation}
f_n(y)=A_n [\cos(m_n y)-\frac{a_0 m_n}{2} \sin(m_n |y|)],
\end{equation}
where $A_n$ is a normalization constant and $m_n$ satisfies 
the eigenvalue equation
\begin{equation}
(4-a_0 a_\pi m_n^2)\tan(\pi m_n R)+2(a_0+a_\pi)m_n=0.
\end{equation}
The main effects of these two brane kinetic terms are to decouple the
heavy KK modes from the brane (it becomes \textit{less transparent} to
them) and to make the mass of the first KK mode light,
\begin{equation}
m_1^2\sim \frac{2}{\pi R}\frac{(a_0+a_\pi)}{a_0 a_\pi}, \mbox{ for
}a_{0,\pi}\gg R.
\end{equation}
This first light KK mode does not completely decouple from the branes
in the large $a_{0}\sim a_{\pi}$ limit but tends to a finite coupling,
which seems necessary to correct the high energy behaviour of the
zero mode coupling~\cite{Carena:2002me}.

The phenomenological implications of brane kinetic terms strongly
depend on the value of the coeffcients. In the particular case that 
$a_0=a_\pi\gg R$ we have a single light KK excitation with coupling
to the branes equal to that of the zero mode, the remaining modes being heavy
($m_{n\geq 2} \sim (n-1)/R$) and with very suppressed couplings to the
branes. In this way we can get 
the first KK mode
fulfilling
the sequential $Z^\prime$ scenario 
(the case with no mixing corresponds to a Higgs in the bulk).
Then all the experimental bounds on such a $Z^\prime$ apply,
$M_{Z^\prime}>1.5$ TeV~\cite{Hagiwara:pw}. This limit is saturated for
large enough boundary kinetic terms~\cite{Carena:2002me}.

A similar behaviour can be observed in the case of
fermions~\cite{apvs}. Consider
as an example the case in which the only relevant brane operator is a
kinetic term for the even component (which we take to be the
RH one),
\begin{eqnarray}
{\mathcal{L}}
&=&
\int {\mathrm{d}}y\; 
\Big\{ [1+a_0 \delta_0 + a_\pi \delta_\pi]
\bar{\Psi}_R {\mathrm{i}} \not \!\partial \Psi_R
+\bar{\Psi}_L {\mathrm{i}} \not \! \partial \Psi_L
+\bar{\Psi}_R \partial_y \Psi_L
-\bar{\Psi}_L \partial_y \Psi_R
\Big\}.
\end{eqnarray}
The orthonormality and differential equations, which determine the KK
expansion, are in this case
\begin{equation}
\int {\mathrm{d}}y\; [1+a_0 \delta_0+a_\pi \delta_\pi] \frac{f_m^R
  f_n^R}{2\pi R}= \delta_{nm}, 
\end{equation}
\begin{equation}
\int {\mathrm{d}}y\; \frac{f_m^L f_n^L}{2\pi R}= \delta_{nm}, 
\end{equation}
and
\begin{eqnarray}
\partial_y f_n^R &=& m_n 
f_n^L,
\\
\partial_y f_n^L &=& -m_n 
[1+a_0 \delta_0 + a_\pi \delta_\pi ] 
f_n^R,
\end{eqnarray}
respectively. 
The two differential equations can then be iterated to get a second
order differential equation for the even component, obtaining 
the same equation as the one for a gauge boson. 
In this case one also obtains a light KK fermion for large brane
terms, and the allowed region in Fig.~\ref{goniometry}, in
particular the cross-point with $m_Q=478$ GeV and $W^L_{tb}=0.96$, can
be recovered for appropriate (large) values of $a_{0,\pi}$.

\end{document}